\documentclass[a4paper]{jpconf}
\usepackage{graphicx}
\usepackage{graphics}
\usepackage{graphicx}
\usepackage{epsfig}
\usepackage{amssymb}
\usepackage{epstopdf}
\def\eq#1{{Eq.~(\ref{#1})}}

\begin{document}
\title{CGC and initial state effects in Heavy Ion Collisions}

\author{Javier L. Albacete}

\address{Institut de Physique Th\'eorique - CEA/Saclay, 
 91191 Gif-sur-Yvette cedex, France.}

\ead{javier.lopez-albacete@cea.fr}

\begin{abstract}
A brief review of the phenomenological studies in the field of heavy ion collisions based on the Color Glass Condensate theory and, in particular, of those relying in the use of the BK equation including running coupling effects is presented.
\end{abstract}

\section{Introduction}

The main goal of the experimental programs on ultra-relativistic heavy ion collisions at RHIC and the LHC is the production and characterization of the Quark Gluon Plasma (QGP). 
A crucial ingredient for the proper understanding of such collisions is the knowledge of the structure of the wave functions of the colliding nuclei, typically encoded in their parton distributions. At high energies, nuclei do not behave as a mere incoherent superposition of their constituents nucleons. Rather, coherence effects are  important and modify not only their partonic content, but also the underlying dynamics of particle production in scattering processes. A careful distinction of such {\it initial state} effects from those originating from the presence of a QGP, or {\it final state} effects, is of vital importance for a proper characterization of the matter produced in heavy ion collisions, as they may lead sometimes to qualitatively similar phenomena in observables of interest.

The Color Glass Condensate effective theory provides a consistent framework to study QCD scattering at high energies (for a review see e.g. \cite{Gelis:2010nm,Weigert:2005us}). The main physical ingredient in the CGC is the inclusion of unitarity effects through the proper consideration of non-linear {\it recombination} effects, both at the level of particle production and also in the quantum evolution of hadronic wave functions. Such effects are expected to be relevant when nuclei (or hadrons, in full generality) are proven at small enough values of Bjorken-$x$. In that regime gluon occupation numbers are very large and gluon self-interactions become highly probable, thus taming, or {\it saturating}, further growth of the gluon densities. 
While the need for unitarity effects comprised in the CGC is, at a theoretical level, clear, the real challenge from a phenomenological point of view is to assess to what extent they are present in available data. Such is a difficult task, since different physical mechanisms concur in data, and also because the limit of asymptotically high energy in which the CGC formalism is developed may not be realized in current experiments. In that sense, the calculation of higher order corrections to the CGC formalism has supposed important leap forward in sharpening the CGC as an useful phenomenological tool.

\section{Some recent developments in the CGC: Running coupling BK equation.}

The leading order BK-JIMWLK equations resums soft gluon emission in the leading logarithmic (LL) approximation in $\alpha_s \ln 1/x$ to all orders, besides of including non-linear terms required by unitarity. At such degree of accuracy, the theory is incompatible with data. Indeed, the $x$-dependence of the saturation scale emerging from the LL equations is $Q_s^2(x)\approx Q_0^2 (x_0/x)^{\lambda^{LL}}$, with $\lambda^{LL} \sim 4.8\alpha_s$ \cite{Albacete:2004gw}, while phenomenological studies of electron-proton scattering and heavy ion collisions indicate that an evolution speed $\lambda\sim 0.2\div 0.3$ is preferred by data. Such insufficiency of the theory has been partially fixed by the calculation of running coupling corrections to the BK-JIMWLK equations through the inclusion of quark loops to all orders \cite{Kovchegov:2006vj,Balitsky:2006wa}. Among other interesting dynamical effects, running coupling effects tame the growth of the saturation scale down to values compatible with experimental data \cite{Albacete:2007yr}. 
Due to the complexity of the JIMWLK equations, in phenomenological works it is more feasible to solve the BK equation, more tractable numerically, which corresponds to their large-$N_c$ limit. It reads
\begin{eqnarray}
  \frac{\partial {\cal N}(r,Y)}{\partial Y}=\int d^2{\bf r_1}\,
  K^{{\rm run}}({\bf r},{\bf r_1},{\bf r_2}) \left[{\cal N}(r_1,Y)+{\cal N}(r_2,Y)-{\cal N}(r,Y)-
    {\cal N}(r_1,Y)\,{\cal N}(r_2,Y)\right]\,,
\label{bk1}
\end{eqnarray}
where $\mathcal{N}(r,Y)$ is the dipole scattering amplitude on a dense target, $Y=\ln x_0/x$ the rapidity and $r$ the dipole transverse size. It turns out that running coupling effects can be incorporated to the evolution equation through just a modification of the evolution kernel, referred to as $K^{run}$ in \eq{bk1}. As discussed in \cite{Albacete:2007yr}, there are several ways of doing so, since the separation of running coupling and conformal terms is not uniquely defined. The evolution kernel proposed in \cite{Balitsky:2006wa} minimizes the role of higher order corrections, making it better suited for phenomenological applications (see \cite{Albacete:2007yr} for an extended discussion on the subject). Finally, \eq{bk1} needs to be suplemented with initial conditions, which can be  choosen to be of the McLerran-Venugopalan type \cite{McLerran:1997fk}. This introduces two free parameters: The value $x_0$ where the evolution starts and the initial saturation scale $Q_0$. Finally, the unintegrated gluon distribution entering the different production processes discussed below is related to the dipole amplitude in \eq{bk1} through a Fourier transform (see \eq{ft}). Besides of the works in heavy ions described below, the rcBK equation has been shown to successfully account for the $x$-dependence of structure functions measured in electron-proton scattering \cite{Albacete:2009fh}.

\section{Total multiplicities in heavy ion colliisions}

The smaller than expected total multiplicities observed at RHIC suggest strong coherence effects in the mechanism for particle production in heavy ion collisions at high energies. The CGC offers a natural explanation of this observation: The total flux of scattering centers (gluons) entering the collisions is significantly reduced due to saturation effects in the wave function of the colliding nuclei. Such idea is realized in the phenomenological KLN model \cite{Kharzeev:2004if}, which relies in the use of $k_t$-factorization. There the total multiplicities in central collisions rise proportional to the saturation scale of the colliding nuclei, $dN/d\eta\propto Q_{sA}^2$. Even though the use of $k_t$-factorization in A+A collisions is not justified, the good description of the energy, rapidity and centrality of multiplicity data yielded by the model lends support to the underlying physical picture. An important improvement of the KLN model was presented in \cite{Albacete:2007sm}, where the phenomenological model for the unintegrated gluon distributions was replaced by exact solutions of the running ccoupling BK equation, also getting a very good description of RHIC data, as seen in Fig. 1. Besides of reducing the degree of modeling needed to describe data, the use of well controlled QCD tools as the rcBK equation allows for a better controlled extrapolation to LHC energies. Taking the centrality class $0-6 \%$ at RHIC as a baseline, the following mid-rapidity multiplicities at the two expected LHC energies in most central Pb+Pb collisions were obtained in \cite{Albacete:2007sm}: 
\begin{eqnarray}
\left. \frac{dN^{ch}_{Pb+Pb}}{d\eta}\right|_{\eta=0}(\sqrt{s_{NN}}=5.5\, {TeV})=1390\pm 95\,;\quad \left. \frac{dN^{ch}_{Pb+Pb}}{d\eta}\right|_{\eta=0}(\sqrt{s_{NN}}=2.76\, {TeV})=1175\pm 75\,.
\label{pred}
\end{eqnarray}  
The error bands in \eq{pred} originate mostly from the uncertainties in the initial conditions for the rcBK equation.
\begin{figure}[t]
\begin{center}
\includegraphics[height=5.5cm]{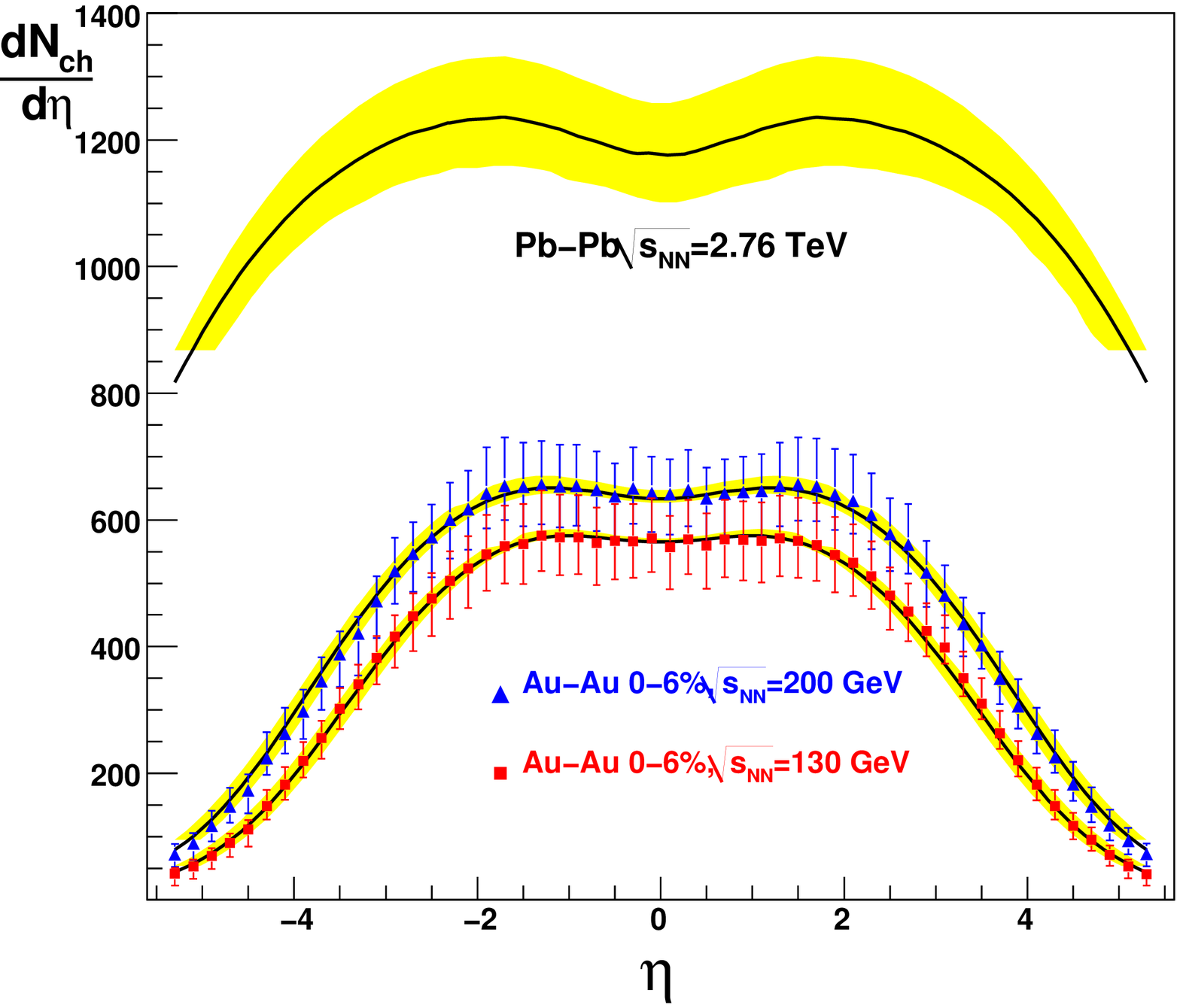}
\includegraphics[height=4.5cm]{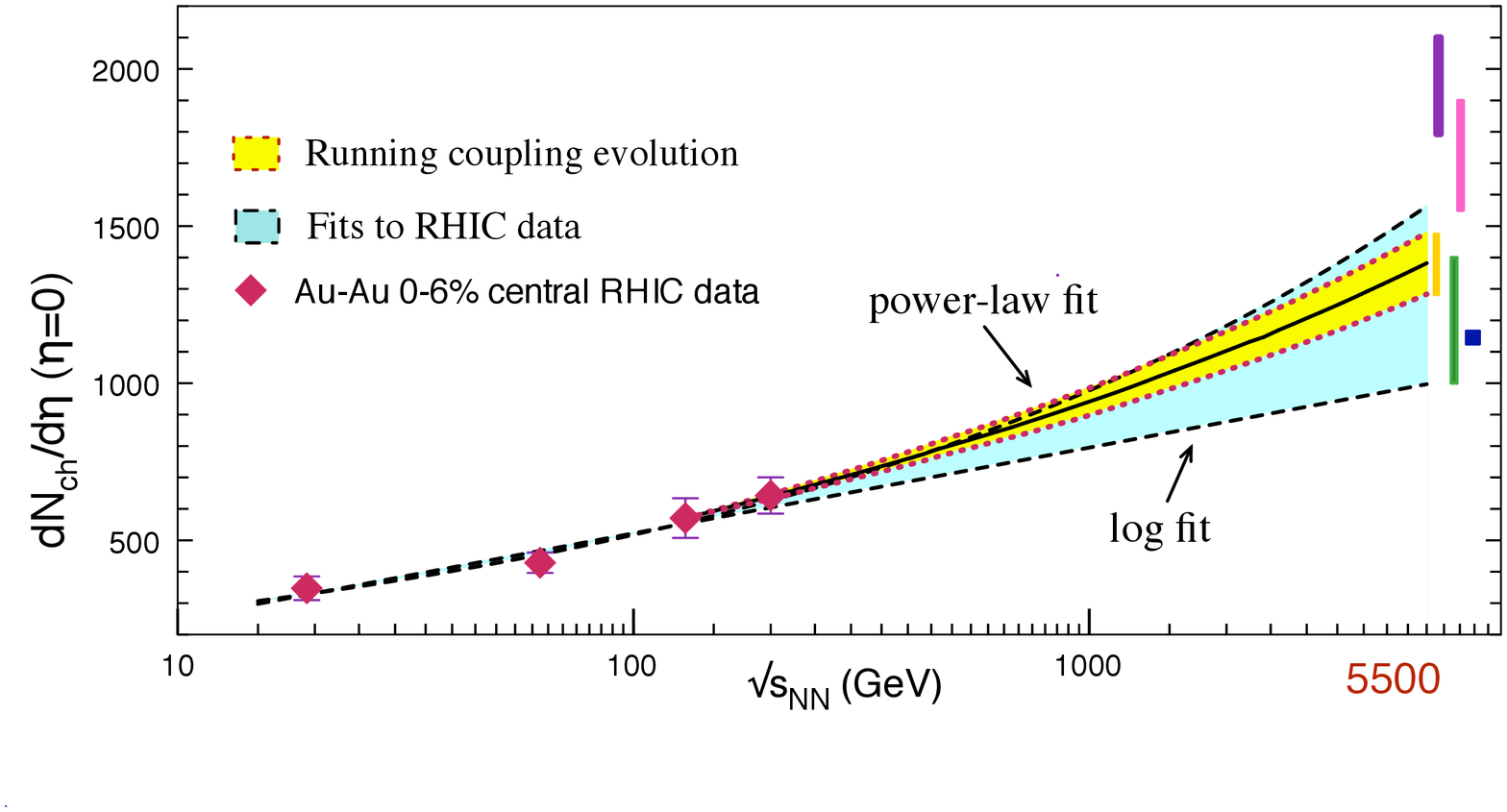}
\caption{Charged particle multiplicity densities in Au+Au collisions at RHIC (data from \cite{Back:2004je}) and predictions for Pb+Pb collisions at $\sqrt{s_NN}=2.76\, TeV$ (left). Mid-rapidity charged multiplicity as a function of the collision energy (left).}
\end{center}
\label{multi}
\end{figure}

\section{Single inclusive particle production and nuclear modification factors}

Particle spectra in A+A collisions are expected to be strongly modified due to the presence of a QGP. Thus, proton (deuteron) nucleus collisions, where no QGP formation is expected, are better suited for the exploration of initial state effects. For any production process, nuclear effects are typically evaluated in terms of the nuclear modification factors:
\begin{equation}
R_{pA}=\frac{\frac{dN^{pA}}{dyd^2p_t}}{N_{coll}\frac{dN^{pp}}{dyd^2p_t}}\,,
\end{equation} 
where $N_{coll}$ is the number of collisions. If high-energy nuclear reactions were a mere  incoherent superposition of nucleon-nucleon collisions, then the observed $R_{pA}$ should be equal to unity.
However, RHIC measurements in d+Au collisions (or peripheral Au+Au collisions) \cite{Arsene:2004ux,Adams:2006uz} feature two distinct regimes: At mid rapidities the nuclear modification factors exhibit an enhancement in particle production at intermediate momenta $p_t\sim 2\div 4$ GeV. In turn, a suppression at smaller momenta is observed. However, at more forward rapidities such Cronin enhancement disappears, turning into an almost homogeneous suppression for all the measured values of $p_t$. According to $2\to1$ kinematics, the $x$-values probed in the projectile and target are $x_{1(2)}=(m_t/\sqrt{s})\,e^{\pm y}$. Thus, for RHIC energies, mid rapidity implies moderate $x\sim 0.1\div 0.01$, probably not small enough to justify the applicability of the CGC. Indeed, mid-rapidity data has been analyzed  through different formalisms and techniques, from collinear perturbation theory  \cite{Eskola:2009uj} to independent multiple scatterings \cite{Accardi:2003jh} or CGC approaches \cite{Kharzeev:2004yx}, all of them reaching a reasonably good agreement with data. 
Data collected in the deuteron fragmentation region offer a cleaner opportunity to explore CGC effects, since the nucleus is proben at muchsmaller values of $x$. There, the CGC formulation of single particle production takes on a relatively simple form \cite{Dumitru:2005gt}: 
\begin{eqnarray}
\frac{dN_h}{dy_h\,d^2p_t}=\frac{K}{(2\pi)^2}\sum_{q}\int_{x_F}^1\,\frac{dz}{z^2}\, \left[x_1f_{q\,/\,p}
(x_1,p_t^2)\,\tilde{N}_F\left(x_2,\frac{p_t}{z}\right)\,D_{h\,/\,q}(z,p_t^2)\right.\nonumber\\ 
+\left. x_1f_{g\,/\,p}(x_1,p_t^2)\,\tilde{N}_A\left(x_2,\frac{p_t}{z}\right)\,D_{h\,/\,g}(z,p_t^2)\right]
\label{hyb}\,,
\end{eqnarray}
where $p_t$ and $y_h$ are the transverse momentum and rapidity of the produced hadron, and
$f_{i/p}$ and $D_{h/i}$ refer to the parton distribution function of the incoming proton and to the final-state hadron fragmentation function respectively. Thus, in the forward region the projectile is in the dilute regime and characterized by its parton distribution functions, while the nucleus is deep in the saturation region and characterized by unintegrated gluon distributions taken from the solutions of the rcBK equation:
\begin{equation}
\tilde{N}_{F(A)}(x,k)=\int d^2{\bf r}\,e^{-i{\bf k}\cdot{\bf r}}\left[1-\mathcal{N}_{F(A)}(r,Y\!=\!\ln(x_0/x))\right],
\label{ft}
\end{equation}
where $k$ refers to transverse momentum.
With this set up we reach a very good description of forward neutral pions and negatively charged hadrons yields as measured by the STAR and BRAHMS Collaborations  respectively in d+Au minimum bias and in p+p collisions, as shown in Fig 2. All the details of the calculation and fit parameters can be found in \cite{Albacete:2010bs}. 
\begin{figure}[t]
\begin{center}
\includegraphics[height=4.5cm]{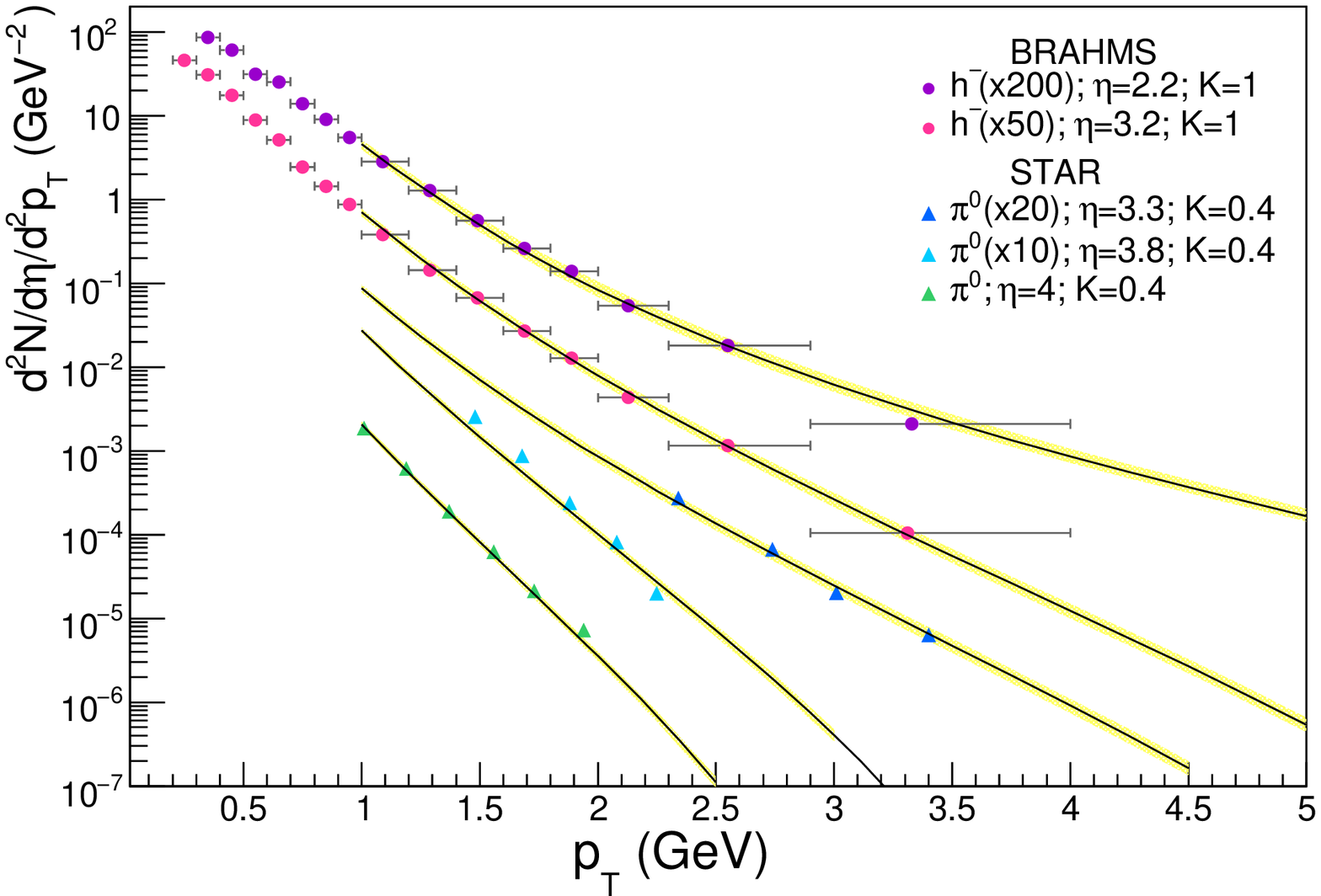}
\includegraphics[height=4.5cm]{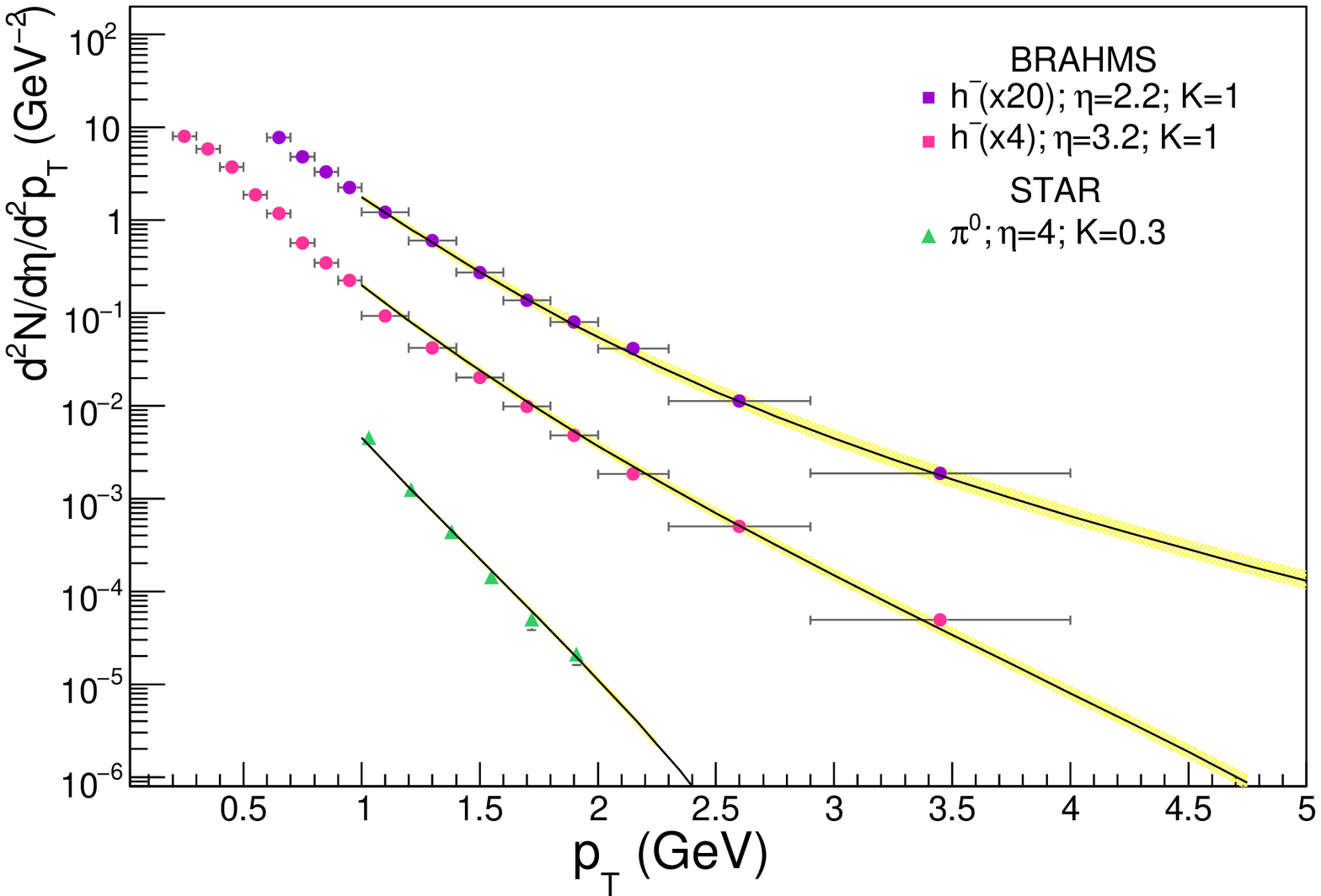}
\end{center}
\caption{ Negative charged hadron and neutral pions at forward rapidities measured by the BRAHMS \cite{Arsene:2004ux} and STAR \cite {Adams:2006uz} Collaborations in p+p (left) and d+Au minimum bias (right) collisions compared to our calculation \cite{Albacete:2010bs}.}
\label{forward}
\end{figure}

By simply taking the ratios of the corresponding spectra, we get a very good description of the nuclear modification factors at forward rapidities. It should be noted that we use the same normalization as the experimentalist do in their analyses of minimum bias d+Au collisions, i.e. we fix $N_{coll}=7.2$. Physically, the observed suppression is due to the relative enhancement of non-linear terms in the small-$x$ evolution of the nuclear wave function with respect to that of a proton.
However, it has been argued that the observed suppression at forward rapidities is not an effect associated to the small values proven in the nuclear wave function but, rather, to energy-momentum conservation corrections relevant for $x_F\to 1$ 
Such corrections are not present in the CGC, built upon the eikonal approximation. Thus, the energy degradation of the projectile parton through either elastic scattering or induced  gluon brehmstralung would be larger in a nucleus than in proton on account of the stronger color fields of the former, resulting in the relative suppression observed in data. A successful description of forward ratios based on the energy loss calculation was presented in \cite{Kopeliovich:2005ym}. 
\begin{figure}[t]
\begin{center}
\includegraphics[height=4.5cm]{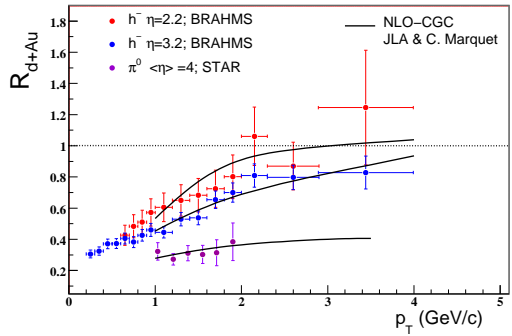}
\caption{Nuclear modification factors at forward rapidities in minimum bias d+Au collisions in the CGC \cite{Albacete:2010bs}. }
\label{ratios}
\end{center}
\end{figure}

Thus, at the level of only this observable, one is in an inconclusive situation. Disentangling the underlying dynamics of the suppression requires either data at larger energies or, alternatively, the study of more exclusive observables. The huge leap forward in collision energy reached at the LHC  allows for an exploration of small-$x$ effects already at mid-rapidity. There, both the target and projectile $x$ are small, and energy loss effects associated to large-$x_F$ effects are expected to be small. In Fig 4. we present our CGC predictions for the nuclear modification factor for negative charged hadrons in p+Pb collisions at two LHC energies. Our curves correspond to rapidities $y=2$ and larger. Technical difficulties related to the intrinsic asymmetry of the formalism used for particle production prevent us from calculating the ratios at mid-rapidity. However, the smooth rapidity dependence suggests that a large suppression $\sim 0.6$ is also expected at mid-rapidity in the LHC. It should be taken into account that the normalization taken to produce the curves in Fig was $N_{coll}=3.6$.   

\begin{figure}[t]
\begin{center}
\includegraphics[height=4cm]{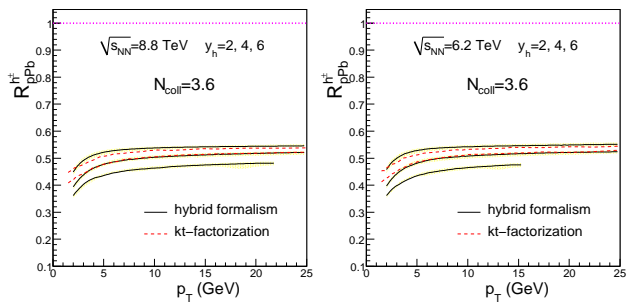}
\caption{CGC predictions for the nuclear modification factor in p+Pb collisions at two LHC energies and rapidities $y_h=2,4$ and 6.}
\label{ratios}
\end{center}
\end{figure}

\section{Forward di-hadron correlations}

As said before, disentangling CGC from other dynamical effects demands the analysis of more exclusive observables. The opportunity is provided by the recent measurement of forward di-pions correlations in d+Au collisions by the STAR collaboration \cite{Braidot:2010zh}. The experimental data for the coincidence probability $CP(\Delta \phi)$ feature a clear suppression of the away-side correlations with respect to the ones observed p+p collisions. This suppression is naturally explained in the CGC: The two pions experimentally observed originate from a valence quark-gluon system in perfect back-to-back correlation.  The quark-gluon system is put on-shell through the interaction with the nucleus, as a result of which the quark and gluon also acquire a transverse momentum of the order of the saturation scale of the nucleus. When that scale, which marks the onset of non-linear effects, is comparable to the initial transverse momenta of the quark and gluon, their intrinsic angular correlation is washed out. Finally, the outgoing quark and gluon fragment independently into hadrons. A recent calculation CGC \cite{Albacete:2010pg} provides a very good comparison with available data. There, the required information on the nuclear wave function is built, through the use of the gaussian approximation, from the two-point function \eq{ft} constrained by the analyses of single inclusive spectra, making it a parameter-free calculation. This study provides, arguably, the most compelling evidence for the presence of CGC effects in available data.

\begin{figure}[t]
\begin{center}
\includegraphics[height=5.5cm]{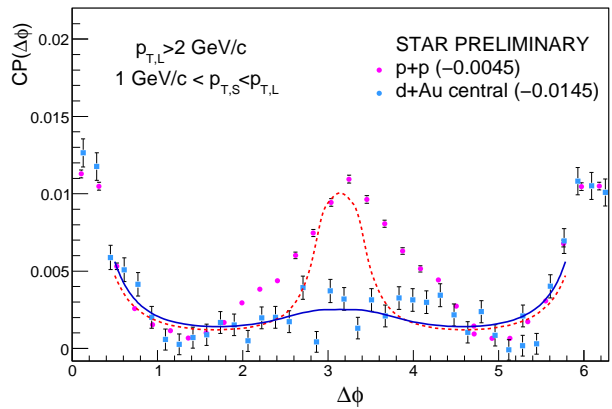}
\caption{Comparison of the CGC calculation of \cite{Albacete:2010pg} with data for the coincidence probability at forward rapidities in d+Au (blue) and p+p collisions (red).}
\label{ratios}
\end{center}
\end{figure}

\section*{Acknowledgments}
I would like to thank the organizers of the Hot Quarks 2010 conference for their invitation to such a nice meeting. I am also grateful to Yuri Kovchegov and Cyrille Marquet, with whom I collaborated in obtaining a good part of the results discussed in the manuscript. This work is supported by a Marie Curie Intra-European Fellowship (FP7- PEOPLE-IEF-2008), contract No. 236376.

\section*{Bibliography}
\bibliographystyle{iopart-num}


\end{document}